\documentclass[10pt,%
               aps,%
               prl,%
               reprint,%
               superscriptaddress,%
               preprintnumbers,%
               amsmath,%
               floatfix,%
               longbibliography,%
               showpacs]{revtex4-1}
\usepackage[T1]{fontenc}
\usepackage[utf8x]{inputenc} 
\usepackage[export]{adjustbox}  
\usepackage[caption=false]{subfig}
\usepackage{url}
\usepackage{color}
\usepackage{float}
\usepackage{amsfonts}
\usepackage{booktabs}  
\usepackage{dcolumn}
\usepackage{wrapfig}
\usepackage{physics}
\usepackage{mathtools}  
\usepackage{enumitem}
\setlist[enumerate]{label=(\roman*), wide}
\graphicspath{{./figures/}}

\usepackage[pdftex,%
            colorlinks=true,%
            linkcolor=blue,%
            citecolor=blue,%
            urlcolor=blue,%
            bookmarksnumbered=true,%
            bookmarksopen=true]{hyperref}
\usepackage[capitalize]{cleveref}

\usepackage{scalerel}
\newcommand{\chis}{\chi^{}_{\scaleto{S}{3.25pt}}}
\newcommand{\chio}{\chi^{}_{\scaleto{0}{3.25pt}}}

\begin{document}

\title{Emergence of a pseudogap in the {BCS}{\textendash}{BEC} crossover}
  

\author{Adam \surname{Richie-Halford}}%
\email{richford@uw.edu}%
\affiliation{%
    Department of Physics,%
    University of Washington, Seattle,%
    Washington 98195--1560, USA
}
  
\author{Joaqu{\'{\i}}n E. \surname{Drut}}%
\email{drut@email.unc.edu}%
\affiliation{%
    Department of Physics and Astronomy,%
    University of North Carolina, Chapel Hill,%
    North Carolina 27599, USA
}

\author{Aurel \surname{Bulgac}}%
\email{bulgac@uw.edu}%
\affiliation{%
    Department of Physics,%
    University of Washington, Seattle,%
    Washington 98195--1560, USA
}
  
\date{\today}

\begin{abstract}
Strongly correlated Fermi systems with pairing interactions become
superfluid below a critical temperature $T_c$. The extent to which such
pairing correlations alter the behavior of the liquid at temperatures
$T > T_c$ is a subtle issue that remains an area of debate, in
particular regarding the appearance of the so-called pseudogap in the
BCS{\textendash}BEC crossover of unpolarized spin-$1/2$ nonrelativistic
matter. To shed light on this, we extract several quantities of crucial
importance at and around the unitary limit, namely: the odd-even
staggering of the total energy, the spin susceptibility, the pairing
correlation function, the condensate fraction, and the critical
temperature $T_c$, using a non-perturbative, constrained-ensemble
quantum Monte Carlo algorithm.
\end{abstract}

\preprint{NT@UW-20-21}

\maketitle

\paragraph{Introduction.}
Dilute, two-component Fermi gases with short-range interactions
are relevant to a variety of systems in nuclear and condensed
matter physics~\cite{ZwergerBook, Strinati_2018}. In ultracold atomic
gases~\cite{RevModPhys.80.885, RevModPhys.80.1215}, the strength
of the interaction can be tuned essentially at will by driving
the system across a Feshbach resonance using an external magnetic
field~\cite{RevModPhys.82.1225}, from a weakly coupled
state, well-described by Bardeen, Cooper, Schrieffer (BCS) theory, to
a state with molecular bound states corresponding to a Bose-Einstein
Condensate (BEC). A smooth crossover~\cite{chen2005bcs,ZwergerBook}
links these limiting regimes as one changes the sign of the inverse scattering
length $1/(k_F a)$, where $k_F$ is the Fermi momentum. On the BCS
side, when $1 / (k_F a) \ll -1$, pairing correlations and Cooper pairs
disappear with the superconducting order parameter $\Delta$ at the
critical temperature $T_c$. Conversely, the BEC regime, where $1 / (k_F
a) \gg 1$, is characterized by the pre-formation of pairs below $T^* \gg
T_c$. It is common to define $T^*$ as the temperature at which pairing
correlations vanish and declare that $T^* = T_c$ on the BCS side.
Between these extremes there exists a ``pseudogap'' regime, where one
finds effects of pairing correlations {\it without} superfluidity and
long range order for temperatures $T_c \le T \le T^*$. The precise
scattering length at which the pseudogap regime begins is still
debated~\cite{mueller2017review}. Specifically, the existence of a
pseudogap in the unitary limit, where $1/(k_F a) = 0$, is not settled.
%

Though the pseudogap is commonly defined as a suppression of the
single-particle density of states near the Fermi surface, there are
several competing definitions, whose differing signatures have led to
debates about their respective existence~\cite{mueller2017review}. The
pseudogap should be identifiable from measurement of the single-particle
spectrum, spin-susceptibility, and even-odd energy staggering,
among others. Even when researchers agree on the definition and
observable signature, there are still subjective judgements regarding
the size of the effects. For example, how much suppression of the
spin susceptibility, or how much even-odd energy staggering above
$T_c$, is necessary to claim evidence for a pseudogap. As argued by
Mueller~\cite{mueller2017review}, the main challenge in understanding and
even defining the pseudogap is that one is dealing with a strongly
correlated system in the normal phase.
On one hand, said strong correlations preclude perturbative approaches.
On the other hand, the lack of order prevents modeling the low-energy
excitations by following the conventional routes of effective field
theory around an ordered state.
(i.e. mean-field or mean-field-plus-fluctuations approaches).
To form a coherent picture of the phenomenology, it is imperative
to continue gathering information on the behavior of these kinds of
systems, in particular the universal, highly malleable ultracold-atom
systems considered here.

We offer perspective on this issue by studying pseudogap signatures
for $0.0 \le 1 / (k_F a) \le 0.3$. We expect to see such signatures
for the highest couplings and then detect either their disappearance
or maintenance as we approach unitarity from the BEC side. We perform
auxiliary-field quantum Monte Carlo (AFQMC) lattice simulations
with constrained ensembles using particle-projection methods, with
a previously introduced model and method~\cite{bulgac2012unitary,
supplement}, modified to employ a cubic (rather than spherical)
momentum cutoff.
As \textcite{werner2012general} explain, a cubic lattice with
an additional spherical cutoff breaks Galilean invariance inducing
$\mathcal{O}(K)$ effects in the effective range expansion, where $K$ is
the center-of-mass momentum of the two-particle system. In contrast, a
pure cubic cutoff breaks the symmetry at $\mathcal{O}(K^2)$.
This is particularly important in the unitary regime where a noticeable
fraction of Cooper pairs have finite $K$.
In addition to previously employed projections
for the total particle number, we introduce a new projection for
the particle asymmetry only, which is free of the infamous sign
problem~\cite{berger2019complex}.
\begin{figure*}
    \subfloat{
        \includegraphics[width=0.58\textwidth,valign=c]{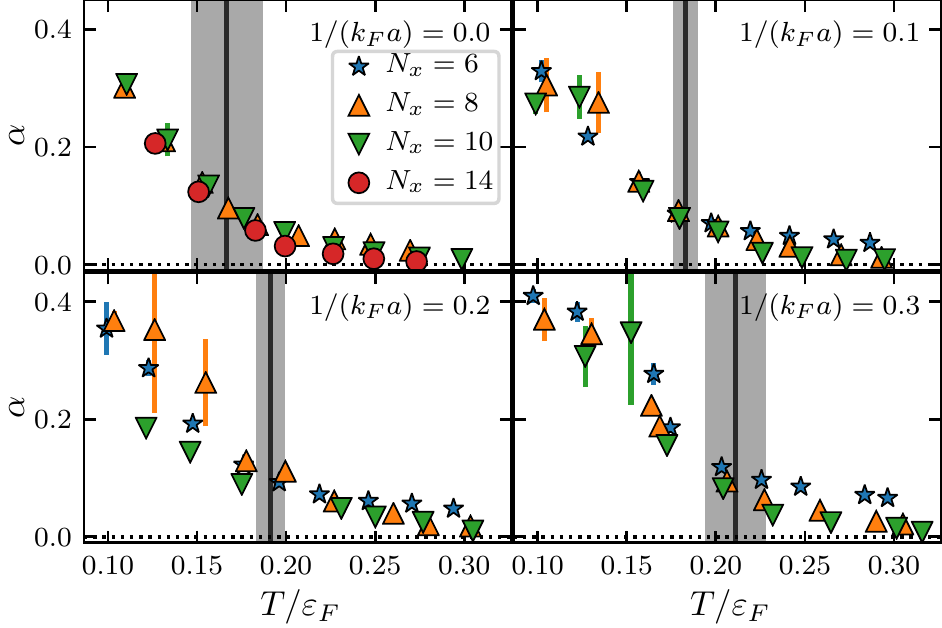}
    }
    \hfill
    \subfloat{
        \includegraphics[width=0.39\textwidth,valign=c]{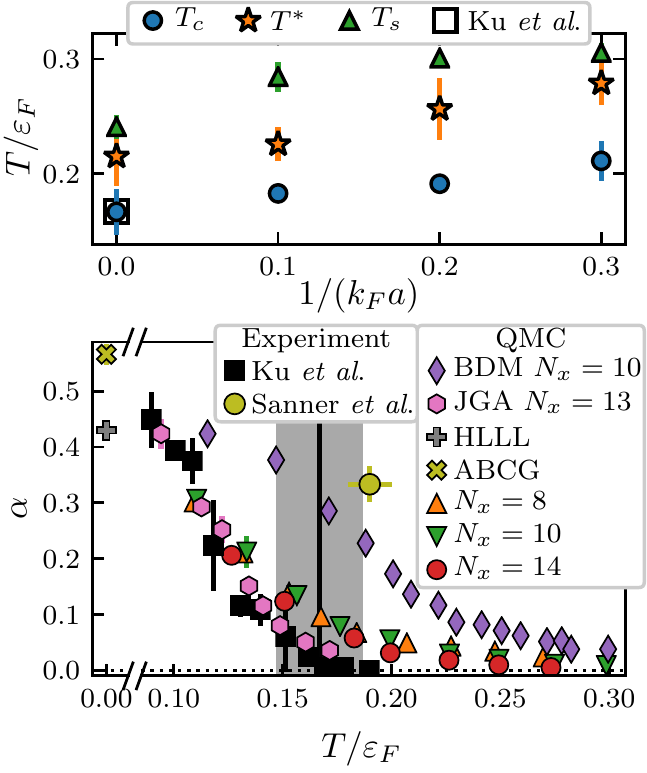}
    }
    \caption{
        Left: The condensate fraction $\alpha$ as a function of temperature 
        at different scattering lengths; at a fixed
        temperature, $\alpha$ increases
        toward the BEC limit.
        At all scattering lengths,
        the condensate fraction tends to decrease with
        an increase in lattice size.
        At $1 / (k_F a) = 0.2$,
        \textcite{astrakharchik2005momentum} estimated the
        zero-temperature condensate fraction as $\alpha(T = 0)
        \approx 0.65$.
        Right (top): characteristic temperatures in the
        {BCS}\textendash{BEC} crossover; $T_c$ is the superfluid
        critical temperature; $T_s$ is a lower bound on the temperature
        at which the spin susceptibility peaks; and $T^*$ is the temperature
        at which the pairing gap disappears. Our estimate for $T_c$ agrees
        with the experimental value from \textcite{ku2012revealing}.
        Right (bottom): $\alpha$ at unitarity;
        error bars for our results are typically within the marker size.
        Also shown: the experimental
        results of \textcite{ku2012revealing}, \textcite{sanner2011speckle},
        and the previous
        AFQMC studies of \textcite{bulgac2008quantum}
        (BDM) and \textcite{jensen2020pairing} (JGA).
        We also plot zero-temperature results by
        \textcite{astrakharchik2005momentum} (ABCG) and
        \textcite{he2019superfluid} (HLLL). 
        The large condensate fraction measured by Sanner \textit{et
        al}.~is relevant to our comparison of the spin susceptibility in
        \cref{fig:chi_all_kfa}.
        The JGA estimates, derived from the maximum eigenvalue of the
        two-body density matrix, are closer to the experimental results
        especially at high temperature, whereas the finite-size scaling
        of our results, derived from the asymptotic values of $h(k_F
        r)$, yields more accurate estimates of the critical temperature
        $T_c$. The discrepancy between our results and BDM, which are
        also derived from the asymptotic behavior of $h(k_F r)$, support
        the argument of \textcite{jensen2020pairing} that the difference
        is due to the BDM spherical momentum cutoff.
        $T_c$ estimates are compatible with previous
        estimates by \textcite{burovski2008critical} and
        \textcite{bulgac2008quantum}. Estimates for $T^*$ are compatible
        with previous results by \textcite{magierski2011onset}.
    }\label{fig:alpha}
\end{figure*}
We simulate on a cubic lattice of
size $L = N_x \ell$, set units such that $\hbar = k_B = m = 1$, and
set the spatial lattice spacing to $\ell = 1$, which is equivalent to a
choice of ``lattice units.'' $N_x$ therefore dictates the lattice size
and approach to the thermodynamic limit. We use $N$ to denote the total
particle number, $N \equiv N_\uparrow + N_\downarrow$, where $N_\sigma$
is the number of spin-$\sigma$ particles with $\sigma \in \{\uparrow,
\downarrow\}$, not to be confused with the particle number asymmetry
$N_{-} \equiv N_\uparrow - N_\downarrow$.

\paragraph{Results.}
We determined the condensate fraction, critical temperature, spin
susceptibility, even-odd pairing gap, and energy per particle.
We also performed the first finite-temperature measurements of
the Tan contact away from unitarity. Given the ongoing debate
over pseudogap signatures and the relationship between the Tan
contact, which is dominated by short-range interaction effects,
and pairing, which characterizes long-range correlations (see
Refs.~\cite{pieri2009enhanced, mueller2017review}), we defer these
results to the supplementary material~\cite{supplement}.
Error bars on individual points represent
statistical errors and show the standard error of the mean. Error bands
in \cref{fig:delta_all,fig:xi_all} incorporate statistical errors and
finite volume effects and represent the standard error of the mean.

\begin{enumerate}
    \item \textbf{Condensate fraction:}
        The condensate fraction can be obtained from the asymptotic
        behavior of the quantity $h(r)$~\cite{astrakharchik2005momentum,
        burovski2008critical, bulgac2008quantum}:

        \begin{align}
            \alpha &= \lim_{r \to \infty} h(r), \quad
            h(r) = \frac{N}{2} \left( g_2(r) - g_1(r)^2 \right), \\
            g_2(r) &= \left( \tfrac{2}{N} \right)^2
            \int \dd[3]{\vb{r}_1} \dd[3]{\vb{r}_2}
            \expval{
                \psi_\uparrow^\dagger (\vb{r}_1')
                \psi_\downarrow^\dagger (\vb{r}_2')
                \psi_\downarrow (\vb{r}_2)
                \psi_\uparrow (\vb{r}_1)
            }, \nonumber \\
            g_1(r) &= \tfrac{2}{N} \int \dd[3]{\vb{r}_1}
            \expval{
                \psi_\uparrow^\dagger (\vb{r}_1')
                \psi_\uparrow (\vb{r}_1)
            },
            \qquad
            \vb{r}_{1, 2}' \equiv \vb{r}_{1, 2} + \vb{r},
            \nonumber
        \end{align}
        which acts as an order parameter, characterizing the extent
        of off-diagonal long-range order~\cite{yang1962concept}.
        In \cref{fig:alpha}, we show our results for $\alpha$ at
        different scattering lengths. An alternative approach
        is to estimate $\alpha$ as the maximum eigenvalue of
        $g_2$~\cite{jensen2019pseudogap}. Comparing our results to those
        of the eigenvalue method, and to experimental values in the
        right panel of \cref{fig:alpha}, suggests that the eigenvalue
        method approaches the experimental $\alpha$ more quickly than
        our asymptotic value method, most noticeably at higher $T$.

        However, we also use the finite-size scaling of $\alpha$
        to determine $T_c$. By calculating $\alpha$ at multiple
        temperatures and lattice sizes, we obtain ``crossing
        temperatures'' (i.e. lattice-size-dependent estimates
        of $T_c$) from which we extrapolate to infinite volume
        to determine the true $T_c$~\cite{bulgac2008quantum,
        astrakharchik2005momentum,burovski2008critical,supplement}. That
        procedure yields $T_c$ as shown in \cref{fig:alpha}, which are
        consistent with previous studies ~\cite{burovski2008critical,
        bulgac2008quantum} and in agreement with the
        experimental result $T_c / \varepsilon_F = 0.167(13)$ at
        unitarity~\cite{ku2012revealing}.

        \begin{figure}[t]
            \includegraphics[width=\columnwidth]{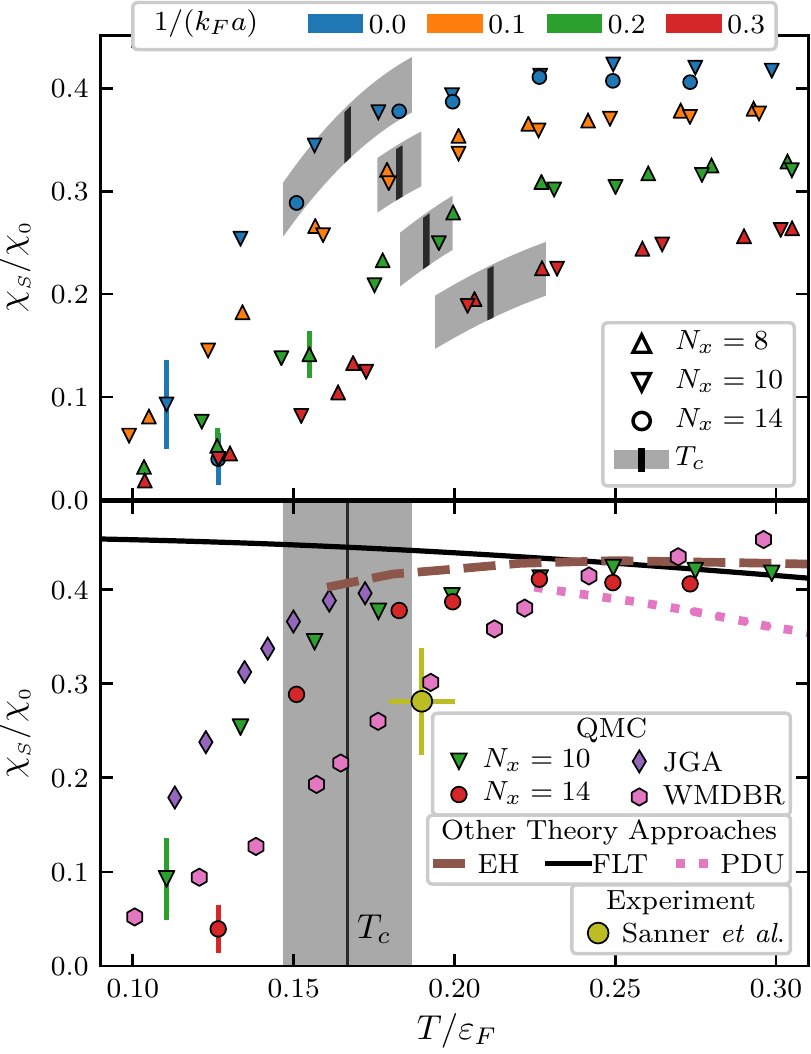}
            \caption{
                Top: AFQMC results for the spin susceptibility $\chis$ at
                four different scattering lengths, scaled by its
                zero-temperature non-interacting counterpart, $\chio
                = 3 N / (2 V \varepsilon_F)$. 
                Bottom: At unitarity, we compare our result to two
                previous AFQMC studies: \textcite{jensen2019pseudogap}
                (JGA) and \textcite{wlazlowski2013cooper} (WMDBR); the
                experimental result of \textcite{sanner2011speckle};
                a self-consistent Luttinger-Ward result (EH,
                \cite{enss2012quantum}); the normal Fermi liquid theory
                prediction; and a self-consistent NSR result (PDU,
                \cite{pantel2014polarized}). 
            }\label{fig:chi_all_kfa}
        \end{figure}
        
    \item \textbf{Spin susceptibility:}
        A probe of the normal state character of the pairing is the
        spin-susceptibility $\chis$, which should be suppressed below
        $T^{*}$, as fermions bind into pairs, making the gas strongly
        diamagnetic~\cite{trivedi1995deviations}. This is also naturally
        related to the fluctuations in particle asymmetry by
        \begin{equation}
            \chis = \frac{1}{TV} \expval{\hat{N}_{-}^2}
            = \frac{1}{TV} \expval{ \left(
                \hat{N}_\uparrow - \hat{N}_\downarrow
            \right)^2 }.
        \end{equation}
        In \cref{fig:chi_all_kfa}, we show our results for $\chis$.
        We use the particle-asymmetry constrained ensemble, which is
        completely sign-problem free~\cite{supplement}. Our results
        demonstrate an expected decrease in the maximal value of
        $\chis$ as $1/(k_F a)$ increases toward the BEC regime. We
        also find a moderate suppression of $\chis$ above $T_c$,
        which increases towards the BEC regime. In the lower panel of
        \cref{fig:chi_all_kfa}, we compare our results at unitarity
        to two previous AFQMC calculations~\cite{jensen2019pseudogap,
        wlazlowski2013cooper}, an estimate using strong-coupling
        Luttinger-Ward theory~\cite{enss2012quantum}, an experimental
        result from \textcite{sanner2011speckle}, the prediction from
        normal Fermi liquid theory (nFLT), and a self-consistent NSR
        estimate from \textcite{pantel2014polarized}. The deviation
        from FLT behavior confirms symmetry based arguments by
        \textcite{rothstein2019symmetry} that 3D unitary Fermi gases
        cannot be adequately described by nFLT in the range $T_c <
        T < T_F$. Our suppression in $\chis$ is less severe than in
        calculations by \textcite{wlazlowski2013cooper}, supporting
        the argument by \textcite{jensen2019pseudogap} that said
        suppression is affected by the choice of spherical cutoff. The
        experimental value is supressed due to their finite condensate
        fraction even above $T_c$, which can be seen in
        \cref{fig:alpha}. However, our spin susceptibility is
        more suppressed than in both Jensen \textit{et al}.~and
        \textcite{enss2012quantum}, and, more importantly, the
        effect seems to grow for larger systems rather than lessen.
        \Cref{fig:chi_all_kfa} also shows our results for the spin
        susceptibility for $0.1 \le 1 / (k_F a) \le 0.3$. To our
        knowledge, these are the first QMC measurements of $\chis$ away
        from unitarity.

        \textcite{tajima2014uniform, tajima2016strong} identified the
        temperature at which $\chis$ peaks as $T_s$, and the temperature
        range $T_c < T < T_s$ as the ``spin-gap'' range where there
        are fewer free spins to contribute to $\chis$. Although they
        find that $T_s \sim T^*$, the exact relationship between these
        two temperatures requires further study. We present only lower
        bounds for the temperature $T_s$ in \cref{fig:alpha}.

    \item \textbf{Energy stagger pairing gap:}
        The even-odd staggering of systems with fixed particle numbers
        has been used as a measure of pairing since early studies
        of nuclear structure~\cite{bohr1998nuclear}. On the other
        hand, the physical origin of the pseudogap, and consequently
        the way one should measure it, has been the core of a long
        debate since the early days of high-Tc superconductivity (see
        \textcite{randeria1997precursor} for a review). It should be
        noted that our use of the even-odd staggering gap as a measure
        of the pseudogap presupposes that the pseudogap origin lies in
        the preformation of Cooper pairs above $T_c$.
        Several finite-difference formulas have been used to circumvent
        this (see Ref~\cite{madland1988new, *moller1992nuclear,
        *duguet2001pairing} for in-depth discussions). The simplest
        one is the three-point estimate, $\Delta_E^{(3)}$, which
        assumes a linear equation of state. If the equation of state
        has positive curvature, $\Delta_E^{(3)}$ will underestimate the
        pairing gap when $N$ is even and overestimate the pairing gap
        when $N$ is odd. Instead, we use the five-point expression
        \begin{equation}
            \Delta_E^{(5)} \!=\! \frac{(-1)^{N}}{8}
            \!
            \displaystyle \sum_{s = \pm 1}
            \!
            \Big[ 
                4E(N \! + \! s) \!
                - \! E(N \! + \! 2s) \!
                - \! 3E(N)
            \Big],
            \label{eq:delta5}
        \end{equation}
        where $E(N)$ is the ground state energy of a system with
        $N$ total particles, which will be achieved when $\abs{N_{-}}=
        \text{mod}(N, 2)$. In addition to calculating $\Delta_E^{(5)}$,
        we propose another estimation method, which is to fit the
        energies calculated for many different values of $N$ and $N_{-}$
        to a two-parameter equation of state,
        \begin{equation}
            \frac{E}{E_{FG}}(\xi, \Delta_E^{(f)})
            = \xi + \abs{N_{-}}
            \frac{\Delta_E^{(f)}}{E_{FG}},
            \label{eq:eosdeltafit}
        \end{equation}
        where $\xi(T/\varepsilon_F, 1/(k_F a))$ is a
        temperature-dependent generalization of the Bertsch parameter,
        $\varepsilon_F = (\hbar^2 k_F^2) / (2m)$ is the Fermi energy,
        $E_{FG} = 3 N \varepsilon_F / 5$ is the energy of a free Fermi
        gas at zero-temperature, and we use $\abs{N_{-}} \in \{0, 1,
        2\}$ for the fitting procedure~\cite{supplement}.
        Regardless of the estimation scheme, we expect $\Delta_E$ to
        become finite below some temperature $T^{*}$. If $T^{*}$ exceeds
        the critical temperature $T_c$, this garners support for the
        existence of a pseudogap.

        \begin{figure}[t]
            \includegraphics[width=\columnwidth]{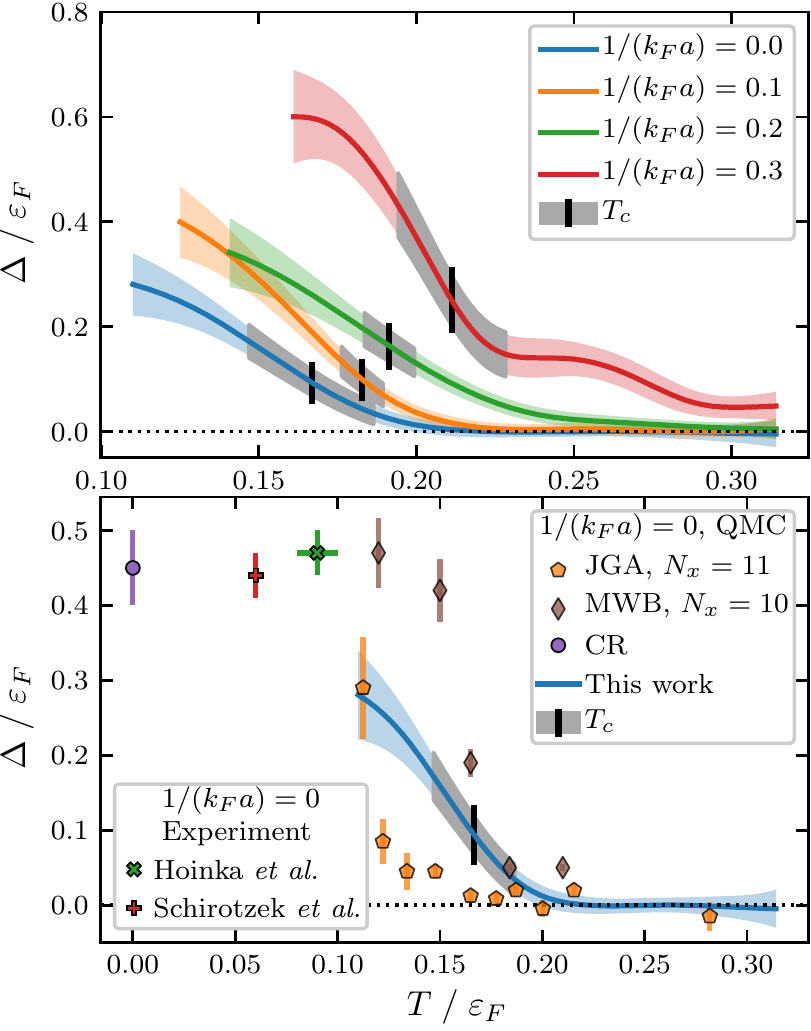}
            \caption{
                Top: AFQMC results for $\Delta_E$ at four different
                scattering lengths, scaled by the Fermi energy
                $\varepsilon_F$. We incorporate results for all lattices
                with $N_x \ge 8$ using a regression technique described
                in the supplementary material \cite{supplement}.
                Bottom: At unitarity, we compare our results to the
                AFQMC results of \textcite{jensen2019pseudogap}
                (JGA) and \textcite{magierski2011onset}
                (MWB); the zero-temperature QMC prediction of
                \textcite{carlson2008superfluid} (CR); and the
                experimental results of \textcite{hoinka2017goldstone}
                and \textcite{schirotzek2008determination}.
            }\label{fig:delta_all}
        \end{figure}

        In \cref{fig:delta_all}, we present our results for the
        even-odd pairing gap, derived from both $\Delta_E^{(5)}$ and
        $\Delta_E^{(f)}$~\cite{supplement}. Our method for calculating
        both the pairing gap and the energy equation of state produces
        a profusion of data points, making visual comparison difficult.
        We therefore plot the results of a regression that includes
        all lattice sizes with $N_x \ge 8$, with further details
        provided in the supplementary material\cite{supplement}. In
        the lower panel, we compare our results at unitarity to previous
        theoretical and experimental studies: an AFQMC measurement
        of the spectral gap which employed a spherical momentum
        cutoff (MWB, \cite{magierski2011onset}); a constrained
        ensemble AFQMC study (JGA, \cite{jensen2019pseudogap})
        that estimated $\Delta^{(3)}$ with a cubic cutoff,
        but without relative temperature corrections, which
        we discuss in the supplement~\cite{supplement}; two
        low-temperature experimental results~\cite{hoinka2017goldstone,
        schirotzek2008determination}; and a zero-temperature QMC
        reference result~\cite{carlson2008superfluid}. We can
        view our results as charting a middle course between the
        Jensen \textit{et al}.~results and the Magierski \textit{et
        al}.~results, all of which can be interpreted as approaching
        the low-temperature reference results. However, the
        comparison is fraught since the spectral gap computed by
        \textcite{magierski2011onset} is \emph{a priori} a different
        quantity than the even-odd pairing gap and the critical
        temperature computed by Jensen \textit{et al}.~is lower than
        ours and also the experimentally determined value.

        Despite the large uncertainties at low temperatures, we can
        appreciate certain features of the pairing gap. It is weaker,
        compared to the low temperature limit, for temperatures above
        $T_c$, however, it cannot be said to vanish immediately above
        the $T_c$ error band even at unitarity. Our estimates for
        $T^*$, derived from spline fits~\cite{supplement} of both
        $\Delta_E^{(5)}$, see \cref{eq:delta5}, and $\Delta_E^{(f)}$,
        see \cref{eq:eosdeltafit}, are presented in \cref{fig:alpha}
        and are comparable with a previous AFQMC study that determined
        $T^*$ from the spectral gap~\cite{magierski2011onset}, as
        opposed to the even-odd energy gap~\cite{jensen2019pseudogap}.
        At $1/(k_F a) \approx 0.3$, we detect a plateau in the pairing gap
        above $T_c$. At this scattering length, $na^3 \sim 1$ so that
        the interparticle separation is of the same scale as the Cooper
        pair size, indicating a crossing into the ``pure'' BEC regime, where the
        pseudogap maintains a plateau to very high temperatures.

    \item \textbf{Energy equation of state:}
        \begin{figure}[t]
            \includegraphics[width=\columnwidth]{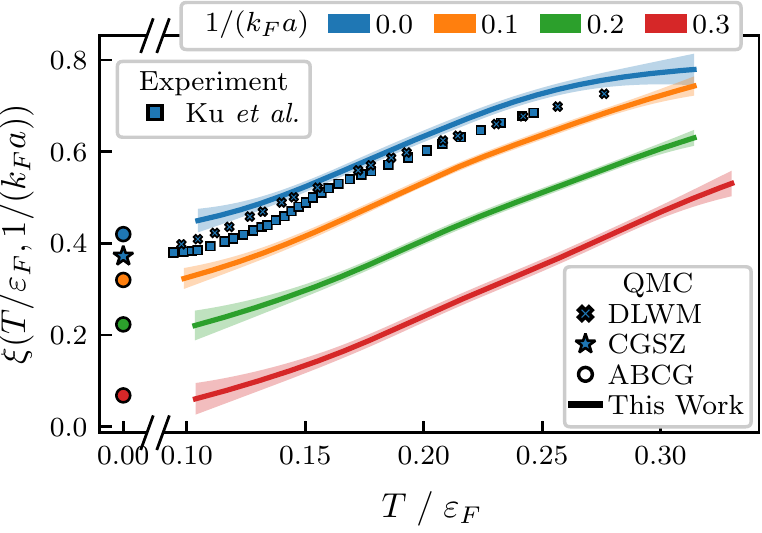}
            \caption{
                AFQMC results for the temperature-dependent Bertsch
                parameter, $\xi(T / \varepsilon_F,1/(k_Fa))$ at
                four different scattering lengths.
                We incorporate results for all lattices
                with $N_x \ge 8$ using a regression technique described
                in the supplementary material \cite{supplement}.
                At unitarity, we compare our results to the
                experimental measurements of \textcite{ku2012revealing}
                and the high-precision AFQMC results of
                \textcite{drut2012equation} (DLWM). We also
                show the zero temperature predictions of
                \textcite{carlson2011auxiliary} (CGSZ) at unitarity and
                of \textcite{astrakharchik2004equation} (ABCG) at all
                scattering lengths.
            }\label{fig:xi_all}
        \end{figure}
        \Cref{eq:eosdeltafit}, which parameterizes the energies of
        systems with various numbers of $N_{\uparrow,\downarrow}$,
        also allows us to extract the temperature- and
        coupling constant-dependent Bertsch parameter $\xi(T/
        \varepsilon_F,1/(k_Fa))$. In \cref{fig:xi_all} we show our
        results for $\xi(T / \varepsilon_F,1/(k_Fa))$ for each
        scattering length and compare to previous results.
        Similar to the results by \textcite{drut2012equation} at
        unitarity, we did not capture the curvature in the equation of
        state seen by \textcite{ku2012revealing} below $T_c$. However,
        our results at unitarity do approach the reference values
        at zero temperature. We have a similar level of agreement with
        the results of \textcite{vanhoucke2012feynman}, which are not
        shown in \cref{fig:xi_all}, but are in excellent agreement with
        experiment in the normal state. We provide a table of values
        and errors for both $\xi$ and $\Delta$ in the supplemental
        material~\cite{supplement}.
\end{enumerate}

\paragraph{Conclusion.}

We performed the first {\it ab initio} finite-temperature calculations
of the spin susceptibility $\chis$ and Tan contact $C$ away from
unitarity, in addition to determining the condensate fraction $\alpha$,
the critical temperature $T_c$, the even-odd pairing gap $\Delta_E$, and
the Bertsch parameter $\xi$.
For both the spin susceptibility and the even-odd pairing gap, we find
no discontinuities as we reduce the coupling, but rather a smooth
reduction in pseudogap signatures.

Since the {BCS}\textendash{BEC} crossover is smooth, we do not
expect an abrupt and discontinuous emergence of the pseudogap.
Questions about where the pseudogap emerges are therefore analogous
to long-debated questions about where the Earth's atmosphere
ends~\cite{goedhart1996never, *mcdowell2018edge}.
Since the field is young, we have not yet developed the pseudogap analog
of the K\'{a}rm\'{a}n line from space science.
We have provided context to this discussion by looking for signatures
of the pseudogap between $0.0 \le 1 / (k_F a) \le 0.3$. At $1 /
(k_F a) = 0.3$, we see strong pseudogap signatures, which diminish
towards unitarity. However, all characteristic temperatures $T^*$ in
\cref{fig:alpha} exceed the critical temperature $T_c$ at all scattering
lengths. Based on our results, we conclude it is premature to exclude
unitarity from the pseudogap regime. Future work should include more
refined extrapolations to the limit of zero-effective range, infinite
volume, and zero density.

\begin{acknowledgments}
{\it Acknowledgments.-}
We thank G.~Wlaz\l{}owski for his valuable input and K.~Roche and S.~Jin
for their guidance on the computational implementation.
ARH and AB were supported by U.S.~Department of Energy, Office of
Science, Grant No.~DE-FG02-97ER41014.
ARH was also supported by the U.S.~Department of Energy, Computational
Science Graduate Fellowship, under Grant No.~DE-FG02-97ER25308.
JD was supported by the U.S.~National Science Foundation under Grant
No.~PHY1452635.
This research used resources of the Oak Ridge Leadership Computing
Facility, which is a US DOE Office of Science User Facility supported
under Contract No.~DE-AC05-00OR22725.
This work was supported by ``High Performance Computing Infrastructure''
in Japan, Project ID: hp180048. A series of simulations were carried out
on the Tsubame 3.0 supercomputer at Tokyo Institute of Technology.
It was also facilitated through the use of advanced computational,
storage, and networking infrastructure provided by the Hyak
supercomputer system and funded by the STF at the University of
Washington.
\end{acknowledgments}


\bibliography{pseudogap}

\end{document}